\documentclass[sigconf]{acmart}

\AtBeginDocument{%
  \providecommand\BibTeX{{%
    \normalfont B\kern-0.5em{\scshape i\kern-0.25em b}\kern-0.8em\TeX}}}

\setcopyright{acmcopyright}

\copyrightyear{2021}
\acmYear{2021}
\setcopyright{acmlicensed}\acmConference[UIST '21]{The 34th Annual ACM Symposium on User Interface Software and Technology}{October 10--14, 2021}{Virtual Event, USA}
\acmBooktitle{The 34th Annual ACM Symposium on User Interface Software and Technology (UIST '21), October 10--14, 2021, Virtual Event, USA}
\acmPrice{15.00}
\acmDOI{10.1145/3472749.3474761}
\acmISBN{978-1-4503-8635-7/21/10}

\usepackage{color}
\usepackage{multirow}
\usepackage{multicol}
\usepackage{booktabs}
\usepackage{enumerate}

\begin{document}

\title{HoloBoard: a Large-format Immersive Teaching Board based on pseudo HoloGraphics}

\author{Jiangtao Gong}
\authornote{Corresponding author}
\affiliation{%
  \institution{Lenovo Research}
  \city{Beijing}
  \country{China}
}
\email{gongjiangtao2@gmail.com}

\author{Teng Han}
\affiliation{%
  \institution{Institute of Software Chinese Academy of Sciences}
  \city{Beijing}
  \country{China}
}
\email{hanteng@iscas.ac.cn}

\author{Siling Gong}
\affiliation{%
    \institution{Lenovo Research}
  \city{Beijing}
  \country{China}}
\email{lily0516gsl@gmail.com}

\author{Jiannan Li}
\affiliation{%
  \institution{University of Toronto}
  \city{Toronto}
  \country{Canada}
}
\email{jiannanli@dgp.toronto.edu}

\author{Siyu Zha}
\affiliation{%
  \institution{Tsinghua University}
  \city{Beijing}
  \country{China}
}
\email{zhasiyu@tsinghua.edu.cn}

\author{Liuxin Zhang}
\affiliation{%
  \institution{Lenovo Research}
  \city{Beijing}
  \country{China}
}
\email{zhanglx2@lenovo.com}

\author{Feng Tian}
\affiliation{%
  \institution{Institute of Software Chinese Academy of Sciences}
  \city{Beijing}
  \country{China}
}
\email{tianfeng@iscas.ac.cn}

\author{Qianying Wang}
\affiliation{%
  \institution{Lenovo Research}
  \city{Beijing}
  \country{China}
}
\email{wangqya@lenovo.com}

\author{Yong Rui}
\affiliation{%
  \institution{Lenovo Research}
  \city{Beijing}
  \country{China}
}
\email{yongrui@lenovo.com}

\renewcommand{\shortauthors}{Gong, et al.}

\begin{abstract}
   In this paper, we present HoloBoard, an interactive large-format pseudo-holographic display system for lecture based classes. With its unique properties of immersive visual display and transparent screen, we designed and implemented a rich set of novel interaction techniques like immersive presentation, role-play, and lecturing behind the scene that are potentially valuable for lecturing in class. We conducted a controlled experimental study to compare a HoloBoard class with a normal class through measuring students’ learning outcomes and three dimensions of engagement (i.e., behavioral, emotional, and cognitive engagement). We used pre-/post- knowledge tests and multimodal learning analytics to measure students’ learning outcomes and learning experiences.
   Results indicated that the lecture-based class utilizing HoloBoard lead to slightly better learning outcomes and a significantly higher level of student engagement. 
   Given the results, we discussed the impact of HoloBoard as an immersive media in the classroom setting and suggest several design implications for deploying HoloBoard in immersive teaching practices.
\end{abstract}



\begin{teaserfigure}
  \includegraphics[width=\textwidth]{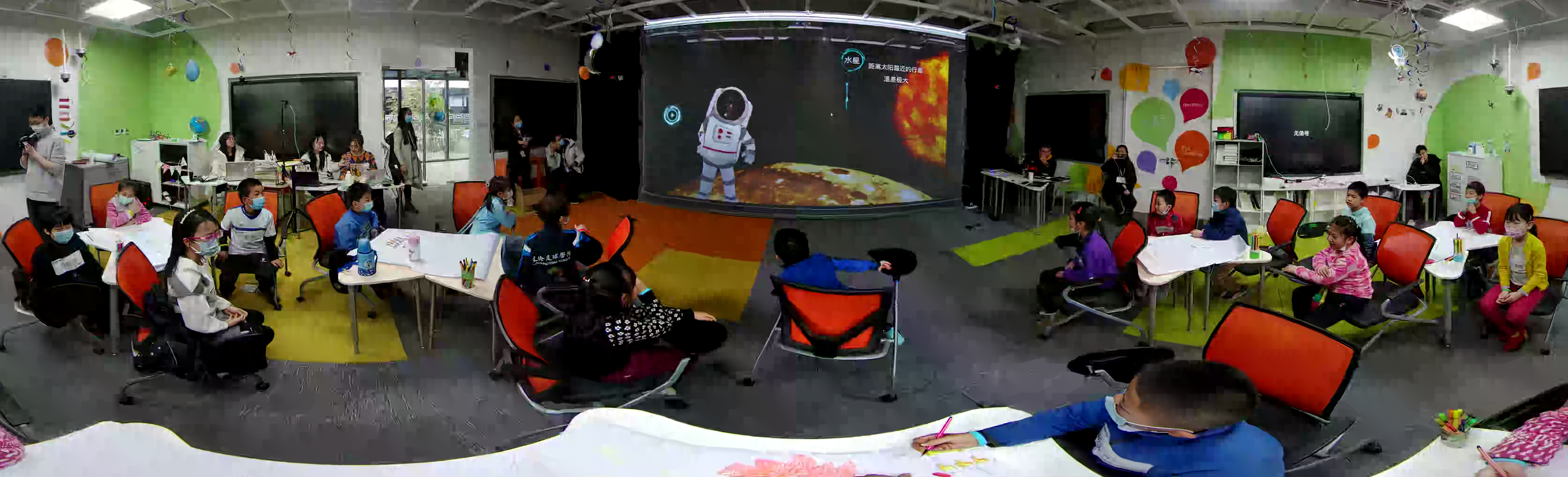}
  \caption{HoloBoard in a regular lecture-based class (recorded by an 8K Panoramic Camera)}
  \Description{Lot of students in a regular classroom with HoloBoard.}
  \label{fig:teaser}
\end{teaserfigure}

\maketitle

\section{Introduction}

Recent advances in understanding new instructional technologies (e.g., interactive whiteboard, VR, AR) indicate a need to rethink how to incorporate active learning into traditional lecturing ~\cite{balta2015attitudes,karsenti2016interactive,ibrahim2018arbis,vazquez2017serendipitous,holstein2018classroom,aslan2019investigating}. With rapid growing and diverse applications, emerging displaying and interaction technologies have seen potential to scale in classroom adoptions.
Indeed, their role in education is attracting increasing attention and is continuously verified in various contexts, e.g., virtual experiment simulation \cite{apostolellis2014evaluating,ferracani2014natural,freina2015literature},  AR-based word learning \cite{ibrahim2018arbis,vazquez2017serendipitous}. However, traditional immersive hardware such as Head Mounted Displays (HMDs) are not entirely suitable for the promoting and applying AR/VR in lecture-based classroom environments. These hardware might cause visual and body discomfort while wearing them, difficulties in facilitating collaborations, dangers of collision during multiplayer mode, and the costly operation/maintenance of the headsets \cite{radu2014augmented}. 

This paper paid attention to teaching boards, which have been used as essential venues to display curriculum materials and help teachers communicate with a large audience in classrooms. Teaching boards have been evolving towards being more functional and interactive in enhancing pedagogical performance \cite{muttappallymyalil2016evolution}. In this work, We present a novel concept of HoloBoard, a pseudo holographic based immersive teaching board system with a large-format transparent display. Pseudo holographics are of interest in science, retail and exhibition applications in recent years. It offers a pseudo 3D image to see near its surface by the reflected light of the dedicated projector.
HoloBoard exploits such capabilities and aims to make lecture-based classes more attractive by blurring the boundary between physical classroom and digital learning content, enabling immersive content display. In comparison with AR/VR, HoloBoard is advantageous in that its setup does not require everyone to wear an HMD, and offers rich interaction possibilities such as co-located mixed reality collaboration with face-to-face communication and natural eye-contact. 
Our work to reach the goal is two-fold: 1) interaction designing of HoloBoard, and 2) verifying its effect in engaging students in lectures. 




The unique visual experiences of watching pseudo 3D images empower us to design a set of interaction techniques that are potentially valuable for lecturing in classroom. To do so, we adopted a user-centric design thinking process (e.g., interview, brainstorming) to excavate teachers’ requirements for an immersive teaching board. With the resulting design considerations, we developed the prototype of HoloBoard associated with advanced interaction techniques ranging widely from immersive presentation to augmented role-play to see-through virtual interactions. Next, to understand how the presentation and interaction of HoloBoard could help teaching and engaging students in lectures, and to give insights on practical deployment of HoloBoard, we carried out a second study in two actual classes with HoloBoard and a regular digital whiteboard, respectively. 
We used pre-/post- knowledge tests and multimodal learning analytics techniques to measure students’ learning outcomes and learning experiences.
Results showed that 
HoloBoard classes created a significantly more positive learning atmosphere, lead to significantly higher student engagement, and resulted in slightly better learning outcomes. Given these, we discussed the impact, design implications and practices of such immersive media in classrooms. 



The paper made the following contributions: i) HoloBoard as a novel large-format immersive teaching board in classroom, built upon pseudo holographic system; ii) design of interaction techniques with HoloBoard that are potentially valuable for lecturing and engaging students; iii) a formal experiment that compared HoloBoard with a regular digital whiteboard and measured students’ learning outcomes and learning experiences; iv) insights on deploying HoloBoard in classrooms. 

\section{Related Work}

Our related work include works which inspired the HoloBoard concept such as immersive teaching technologies, interactive technology for learning, in-class engagement, and multimodal learning analytics.

\subsection{Immersive Teaching Technologies}

Lecture-based class leverages the naturalness of spoken communication and creates a social situation that makes both the lecturer and the audience engaged \cite{charlton2006lectures}.
The role of teaching boards is critical in lecturing activities.
The importance of teaching boards in the education domain has been thoroughly discussed, especially in displaying curriculum materials and enhancing pedagogical performance \cite{gran2018tabula}.
Over the past decades, a significant body of research focused on the impact of Interactive Whiteboards (IWBs)~\cite{betcher2009interactive} on learning~\cite{morgan2008improving,yang2012interactive,balta2015attitudes,karsenti2016interactive}, noting their positive effects on student engagement, behavior~\cite{morgan2008improving}, and learning outcomes~\cite{yang2012interactive}. 

Nowadays, immersive displaying technologies are getting prominent in educational applications ~\cite{apostolellis2014evaluating,ferracani2014natural,freina2015literature,ibrahim2018arbis,vazquez2017serendipitous}. 
Virtual reality (VR) technology is entering mainstream consumer markets through products such as HTC Vive and Occulus Rift/Quest. Both educators and learners share positive attitudes towards using VR for educational objectives~\cite{mikropoulos2011educational} and numerous studies have reported the benefits of VR in education, which include improving time-on-task~\cite{huang2010investigating}, enjoyment~\cite{apostolellis2014evaluating,ferracani2014natural}, motivation~\cite{freina2015literature,sharma2013virtual}, and long-term retention~\cite{rizzo2006virtual}. 
Researchers have found that immersive VR has an advantage over the desktop systems when the tasks involved “complex, 3D, and dynamic” content~\cite{mikropoulos2011educational}. 

Augmented reality (AR) technology has made significant advances over recent years. 
Many previous AR systems for education focused on using HMDs, mobile phones, or projectors to enhance learning equipment or environments for students 
in studying a wide range of subjects, such as 
language~\cite{ibrahim2018arbis,vazquez2017serendipitous}, chemistry \cite{fjeld2007tangible,song2011chemo}, and mechanic engineering \cite{ibanez2014experimenting,enyedy2012learning}. 
Some work explored augmenting teachers' views to assist them with classroom routines (e.g. evaluating student's performance~\cite{holstein2018classroom}).
Overall, AR can support pedagogical processes (e.g., providing scaffolding to students) and promote students' engagement~\cite{aslan2019investigating}.

HoloBoard is built upon an alternative displaying technology, namely pseudo holographics, to enrich the interaction vocabularies around digital teaching boards for more substantial engagement.

\subsection{Interactive Technology for Learning}

The human-computer interaction and learning science community have long recognized the potential of interactive technology for increasing engagement, as they vastly broaden the possibilities for students to explore class materials and actively participate in learning processes. Simple student response systems (also know as `clickers'), which offer a small number of buttons for students to provide answers, were found to help students better get involved in large enrollment courses~\cite{Trees2007TheLE}. General purpose computing devices such as tablets support students participating in richer kinds of learning activities, including drawing~\cite{couse2010tablet}, educational games~\cite{furio2013effects}, and simulated labs~\cite{nedungadi2013enhanced}.
Experiments have confirmed their effects in improving engagement~\cite{couse2010tablet} and learning outcomes~\cite{hassler2016tablet}. 
In addition, experimental interfaces such as tangible user interfaces~\cite{xie2008tangibles,schneider2016using} and robots of varying form factors~\cite{ozgur2017cellulo,tanaka2012children} have been applied in classroom settings and shown their benefits for increasing student engagement, enjoyment~\cite{ozgur2017cellulo,xie2008tangibles}, and sometimes learning outcomes~\cite{tanaka2012children,schneider2016using}.

A large body of research focused on encouraging students' participation through whole-body movements~\cite{kang2016sharedphys,lindgren2013emboldened,gallagher2015enactive,kontra2015physical,gelsomini2020embodied}.
These efforts were supported by an embodied perspective of cognition, which asserts that human cognition is affected by the body's interaction with the physical world~\cite{pecher2005grounding,kontra2015physical}. 
Additionally, whole-body interaction promotes student engagement as it offers immersive experiences~\cite{adachi2013human}, facilitates social interaction in classrooms~\cite{kang2016sharedphys}, and possibly transforms learning activities into performances~\cite{gelsomini2020embodied}. 
For example, Kang et al. explored combining immersive displays, whole-body interaction, and physiological sensing for teaching children knowledge about human body~\cite{kang2016sharedphys}.
They found that the system enabled overall high levels of engagement and enjoyment, lively interaction between students, and strong learning potential.

Leveraging the unique affordances of transparent holographic displays, HoloBoard enabled a new array of classroom activities to engage students, such as instructor role playing, large-format virtual demonstration and multi-user game, and double-sided interactions between students/instructors or students/students.

\subsection{In-class Engagement and Multimodal Learning Analytics}
Many aforementioned studies~\cite{adachi2013human,couse2010tablet} have focused on using educational technology to improve student engagement because high levels of engagement can compensate for low academic achievement, negative affective states (e.g., boredom), and high dropout rates among students~\cite{fredricks2004school,gao2020n}.

It is widely accepted that engagement is a multifaceted concept with three dimensions: \emph{behavioral} (active participation and involvement in activities), \emph{emotional} (positive reactions and feelings towards teachers and work), and \emph{cognitive} (efforts and concentration on completing work) engagement~\cite{fredricks2004school,gao2020n,stroud2015learner}. There are various ways to measure engagement in educational related studies. For example, the widely agreed upon student engagement signs for behavior engagement include upright or close posture, gaze and focus on the teacher or the task, and active participation in classroom discussion and they have been captured in the coding schemes used in ~\cite{becsevli2019mar,hsieh2015exploring}. For emotional engagement, despite self-report, students' affective states are commonly used for analysis, which can be based on arousal and valence feature analysis~\cite{di2018unobtrusive,alyuz2016towards}. Classroom discourse analysis, cognitive level of speech, and the interactive nature of conversations have been used for measuring cognitive engagement~\cite{stroud2015learner,smart2013interactions,zhu2006interaction}. More specifically, previous studies in the field of education and human-agent interaction have investigated various discourse attributes, including total speaking turns taken, total words spoken, and mean words per sentences~\cite{serban2017interactive, stroud2015learner,black2009comparison}.

Considering the rich signs of student engagement, we adopted multi-level multimodal learning analytics to analyze the classroom data. Multimodal learning analytics in education is an emerging branch of learning analytics, and it typically analyzes natural synchronized communication modalities, including speech, gestures, facial expressions, gaze, and embodied actions~\cite{oviatt2018ten,bateman2017multimodality,worsley2015leveraging}. Analysis of multimodal data can take place at multiple levels~\cite{oviatt2018ten}. Taking speech as an example, it can be analyzed at the signal level (e.g., frequency and loudness), activity level (e.g., number of words spoken), or other levels~\cite{oviatt2018ten}. 

In addition, Oviatt et al. ~\cite{oviatt2018ten} suggested that such multi-level multimodal learning analytics can support a more comprehensive understanding of the complex learning process, such as the impact associated with implementing a new educational technology. Therefore, we hope to obtain a deep view of the impact of HoloBoard on students' in-class engagement by utilizing multi-level multimodal learning analytics.

\section{HoloBoard}

HoloBoard aims to provide unique and immersive lecturing techniques with large-format pseudo holographics to engage students in learning processes. 
It was developed based on pseudo holographic projection, which was scalable in size and of a formfactor close to typical teaching board. We conducted interviews with 6 professional teachers to understand their teaching activities and requirements, which were then mapped to ten interaction techniques of HoloBoard. We built a working prototype of HoloBoard with the resulting design considerations.

\subsection{Concept Design}

To establish an initial understanding of traditional teaching activities and to better understand how the concept of HoloBoard can be incorporated into classrooms, we conducted a series of semi-structured interviews with 6 school teachers (4 females, average of 5.58 years of teaching experience in English, Math, Physics or Music in primary school, middle school or high school). 
The interviews with the teachers were aimed to understand their teaching process and activities in classroom. Besides, in the interviews, we asked questions regarding their teaching strategies, multimedia equipments and supplementary materials used in class, their preferred teaching methods to handle different types of content, and approaches to interact with students in teaching. 

\begin{figure}[h]
  \centering
  \includegraphics[width=\linewidth]{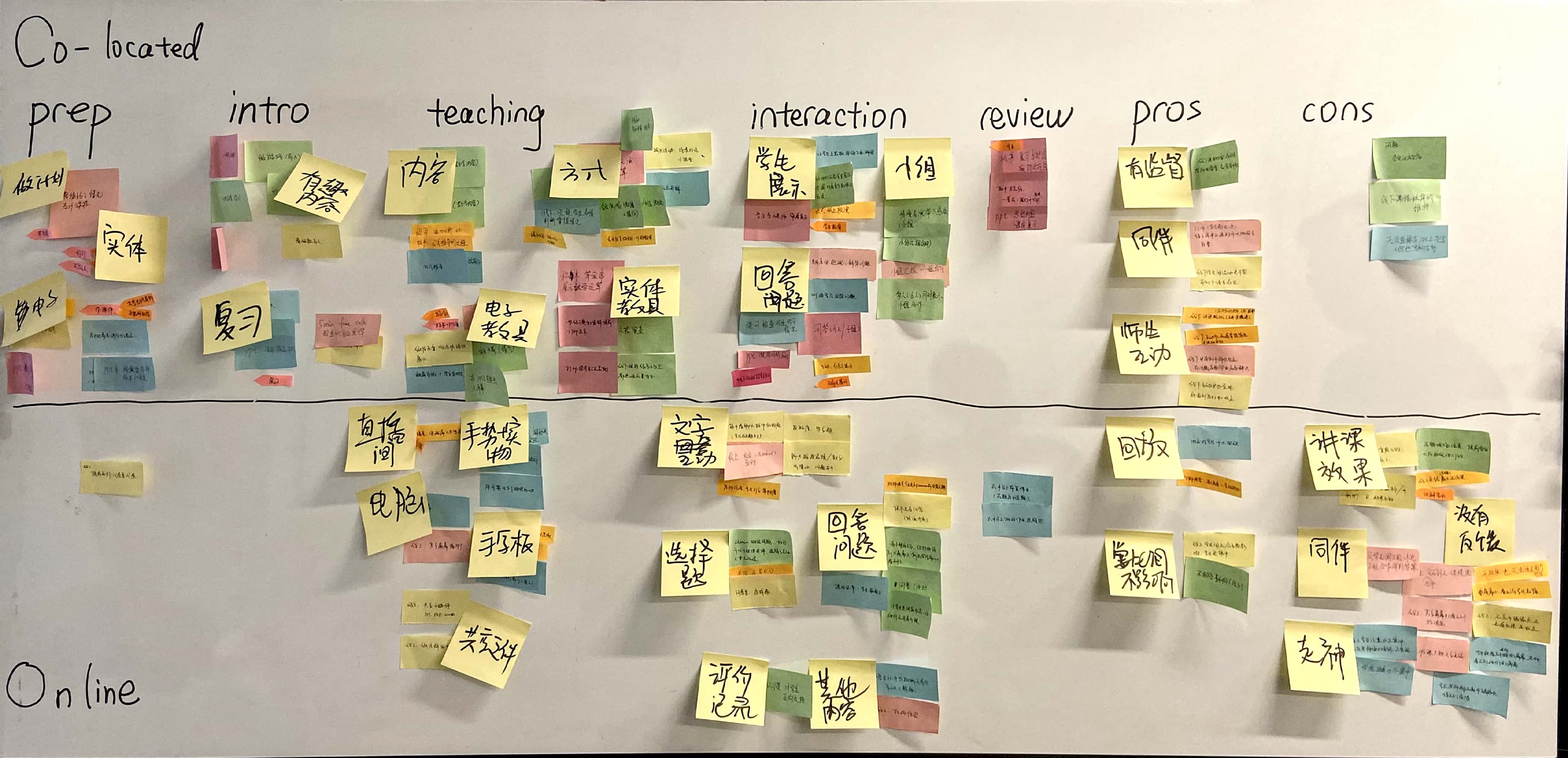}
  \caption{Affinity Diagram of Teaching Activities Interview.}
  \label{affinityDiagram}
\end{figure}

All the interviews were audio-recorded and transcribed. In total, 350 minutes of audio materials were collected. The interview transcripts were analyzed using the open coding method~\cite{corbin2014basics}. Afterward, all codes were transcribed on sticky notes and arranged based on the teaching process via affinity diagramming (see Fig. \ref{affinityDiagram}). 

From the affinity diagram, we obtained the main flow of a typical lesson described as following steps: pre-class preparation (designing the course content and preparing physical and digital materials), warm-up (sharing interesting audio or video clips), lecturing (slides sharing, writing, performing, and physical demonstration), teacher-student interaction (answer questions and students' presentation), and review (write down key takeaways on the blackboard or slide decks). During lecturing, we found that all the teachers have two basic requirements of teaching board: i) to display electronic teaching content of various forms, such as slides, text and video display; ii) to support writing and freehand drawing. 

In addition to basic requirements, we found the teachers have advanced requirements to guide students' imagination and enhance their attention. For instance, the teachers who teach primary school students tend to incorporate body language into teaching: T5 reported that when she taught students arithmetic: ``\textit{sometimes I also design exaggerated performances or teach them some specific actions to remember this formula.}'' T6 used movement to deepen students' understanding of rhythm: ``\textit{I will tell them what kind of scenes they can imagine when they hear this music and then lead them to walk in the classroom following the rhythm of the music.}'' 
Besides, teachers for classes that require demonstrations used tools and proxies such as physics lab equipment (Physics teacher), toy blocks (Math teacher), and musical instruments (Music teacher). These physical demonstration tools also could encourage students to engage, explore, experiment, and learn actively.

\subsection{Interaction Techniques with HoloBoard}


We incorporated the resulting findings from the interviews and the unique presentation properties of pseudo holographics into 10 interaction techniques that are potentially valuable for lecturing scenarios. 



\subsubsection{Immersive Presentation}

The pseudo holographic projection has the ability to create the illusion of 3D objects in volumetric space. The projection space is large, bringing immersive visual experiences to the audience. 
These features give HoloBoard a bunch of novel capabilities for large-format immersive content display (Fig.~\ref{storyboard}a), uniquely suitable for drawing large audiences' attention with immersive visual presentations. This vivid and engaging visual experience is unique to the HoloBoard and can not be achieved with regular digital board.



\subsubsection{Lecturing Behind the Scene} 

\begin{figure*}[h]
  \centering
  \includegraphics[width=0.8\linewidth]{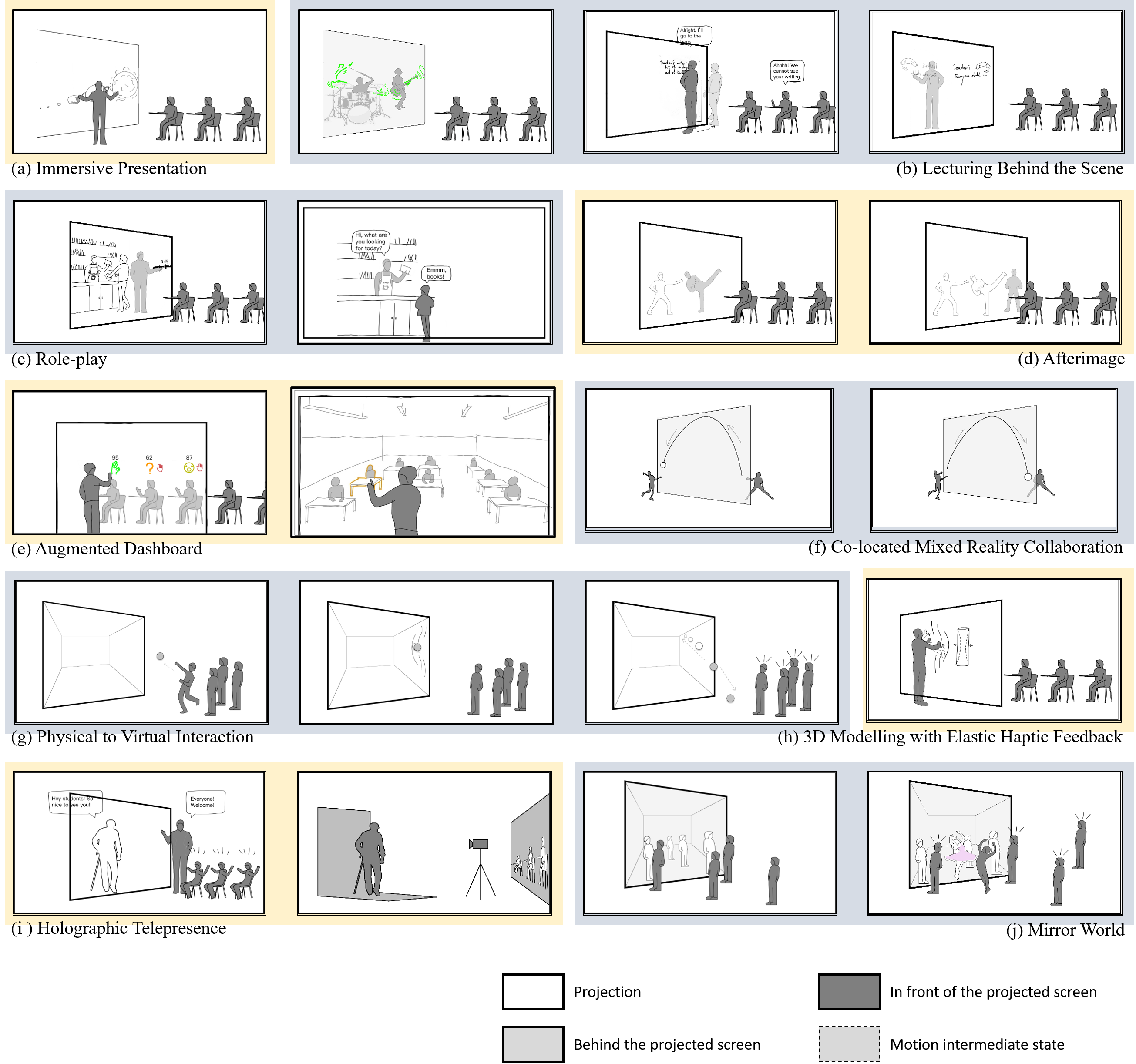}
  \caption{Interaction Techniques: (a) Immersive Presentation: large-format immersive content display; (b)Lecturing Behind the Scene: performing behind the teaching board with real-time special effects overlay and writing on both front and back sides; (c) Role-play: role-playing as a salesperson for situational language exercise; (d) Afterimage: teaching taekwondo and freeze the teacher's current action as an afterimage; (e) Augmented Dashboard: the student's name and relevant information floating around them on the dashboard; (f) Co-located Mixed Reality Collaboration: the user in front the screen can hit the virtual ball towards the backside of HoloBoard; (g) Physical to Virtual Interaction: users can throw a physical ball to the large-format screen. A virtual ball is created when the physical ball hits the screen and the virtual ball moves along the trajectory of the physical one rendering on the HoloBoard; (h) 3D Modelling with Elastic Haptic Feedback: users can perform 3D modeling tasks on the screen, while gaining haptic experiences like hands reaching into the scene;(i) Holographic Telepresence: displaying a remote teacher vividly in the current physical classroom; (j) Mirror World: displaying a digital twin of the physical world on the HoloBoard.}
  \label{storyboard}
\end{figure*}

 HoloBoard allows teachers to lecture behind the teaching board, with digital sketches and writings floating between them and the students. Previous researches \cite{ishii1992clearboard,saquib2019interactive} have demonstrated that real-time graphic overlays do not interfere with the natural conversational style of the presenter, allowing users to effortlessly enhance their communication with the audience, the HoloBoard takes advantage of this flexible and highly adaptable communication style.
One technical challenge posed by the transparent interface are the inverted perspectives of the presenter and the audience. We designed a method of text reversal in order to display the correct text for the presenter and the audience simultaneously (Fig.~\ref{storyboard}b). 
%

\subsubsection{Role-play} 

With the transparent display, students can see pseudo 3D images without losing sight of the presenter behind the HoloBoard. Thus, HoloBoard supports live action role play, during which teachers or students literally act out fictional roles within a narrative (Fig. \ref{storyboard}c). Users behind the board are visually augmented by a digital the avatar overlay, which is animated in real-time according to the users' movements. Users directly drive the generation or motion of graphics by performing specific poses or gestures.
This form of interactive and collaborative storytelling gives an immersive and complex sense of a narrative engagement to students.


\subsubsection{Afterimage} 

HoloBoard integrated techniques of motion visualization in order to convey complicated actions. In specific, the system captures the motion of the presenter, and produces a visualization of the spatiotemporal representations of the presenter as she moves through the space or gestures (as shown in Fig \ref{storyboard}d). 
For instance, when teaching martial arts, taekwondo, dance, etc., the teacher can freeze his current action at any time as an afterimage. In this way, the teacher can give a detailed explanation of these movements enabling students to perform imitation exercises step by step, as necessary.



\subsubsection{Augmented Dashboard} 

Research indicates that teachers are interested in having access to real-time status of students ~\cite{holstein2018classroom,aslan2019investigating}. HoloBoard supports such applications by displaying students' learning analytic information (e.g., test score, emotion) to teachers, undated in real time. 
With HoloBoard, the student's name and other relevant information is associated with each student, enabling teachers to modify lesson plans spontaneously in accord with the changing needs of students, as indicated by various features of the augmented overlay (e.g., changing name tag's color). 
Additionally, the teacher can give interactive markers (e.g., praise) to individual students by pointing to and clicking on the student’s name tag, as shown in Fig. \ref{storyboard}e.

\subsubsection{Co-located Mixed Reality Collaboration}
Due to its large-format and transparency, HoloBoard supports  high quality co-located multi user interactions. HoloBoard also supports co-located users with face-to-face communication. Typically, with a digital whiteboard, two users both have to face the screen, obscuring face-to-face communications, e.g., eye-contact is impossible. HoloBoard overcomes these obstacles of face-to-face communication, allowing users to stand on each side of the board, respectively, and shared a common use space with fully interactive contents. Because users can still see each other clearly, their communication remains natural and spontaneous, enhancing the co-located mixed reality collaboration, as shown in Fig. \ref{storyboard}f. 



\subsubsection{Physical to Virtual Interaction}

In typical classroom, students barely have the chance to interact with the digital contents on screen when they are seated. HoloBoard provides opportunities to interact with the screen in multiple ways. For instance, students in front of the screen can throw a soft ball onto the screen to make a selection or play games. Such method can be further extended by rendering a virtual ball that moves along the trajectory of the real one, thus letting the physical actions reach into the virtual world. Previous researches\cite{jones2013illumiroom,jones2014roomalive} also used surrounding projectors to make physical object interact with virtual content and create an immersive gaming experience in a room space. This helps expanding the interaction space of the students and creating realistic experiences with simulated scenes as shown in Fig. \ref{storyboard}g.


\subsubsection{3D Modelling with Elastic Haptic Feedback}

The projection screen of HoloBoard is made of gauze that has a certain elastic property. Previous work~\cite{watanabe2008deformable,han2014trampoline,muller2014flexiwall} have shown that operations in Z-direction on the screen and the haptic feedback of touching on elastic surfaces helped improving 3D content presentation and maneuvering experience, for instance, improving depth perception. HoloBoard is capable of supporting teachers and students to perform 3D modeling tasks on the screen, while gaining the haptic experiences like hands reaching into the scene as shown in Fig. \ref{storyboard}h.



\subsubsection{Holographic Telepresence}

Holographic telepresence is another scenario that fits the use of HoloBoard \cite{paredes2019my,aman2016exploring,luevano2015use}. This helps extending the face-to-face experience of in-class instruction to remote areas. In this case, it can serve as a medium for remote education and mixed remote panels, displaying remote teachers vividly in the classroom as shown in Fig. \ref{storyboard}i.

\subsubsection{Mirror World}

HoloBoard can serve as a medium for a co-located virtual mirror world, displaying digital twin of the physical world. HoloBoard can elaborate on the learning benefits of the whole-body interaction based on mirror world, which has been shown effective for learning knowledge about the human body~\cite{kang2016sharedphys}, physical experience in science learning~\cite{kontra2015physical}, and other embodied learning. Different from the telepresence where users are remotely located and can not observe the environment and people on site, the mirror world technique allows the teachers and students to see themselves interacting with digital objects and be aware of the surroundings ( Fig. \ref{storyboard}j).

\subsection{Prototype Development}

We built a working prototype of HoloBoard, with which we developed the demo applications enlightened from the aforementioned interaction design.  

\subsubsection{Hardware and Software}
\begin{figure*}[h]
  \centering
  \includegraphics[width=0.8\linewidth]{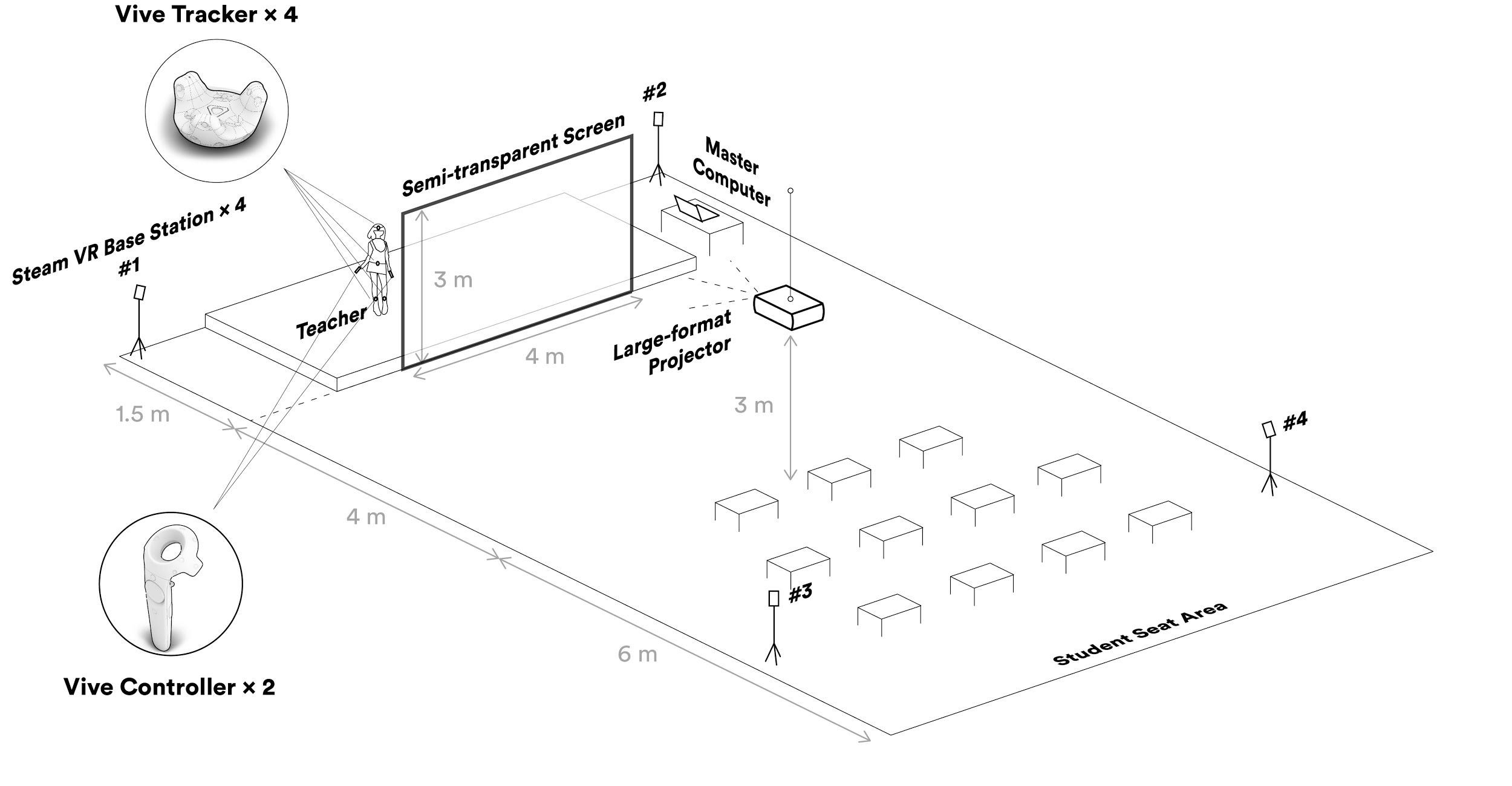}
  \caption{Hardware Composition of HoloBoard}
  \label{system}
\end{figure*}

The system of HoloBoard was developed in a classroom, and integrated the displaying and tracking components (Fig. \ref{system}). In terms of display, we used a 4m×3m large-format projected screen made of translucent (semi-transparent and reflective) material to display digital curriculum material in the front of the classroom. Behind the projected screen is the teacher's stage. Moreover, to further immerse students into the scene, a light-absorbing curtain is hung behind the stage, and 4 adjustable spotlights are put above the staging area to cast more light on the teacher. Thus, the observer can spot the people behind the screen while the projected screen displays. 

In terms of motion tracking, we used 4 SteamVR 2.0 base stations at the 4 corners of the classroom, one on each corner. Two of them were in front of the projection screen while the other two behind the screen. We asked the users to hold the HTC Handler to interact with the HoloGraphics with hands, and wear 4 VIVE trackers on the head, waist, left and right feet, respectively, to track their position and body motions. The system including the communication servers, graphic rendering, and demo applications were implemented with Unity. Many interaction techniques designed for HoloBoard shared the same set of displaying, tracking setup.



 


\begin{figure*}[h]
  \centering
  \includegraphics[width=\linewidth]{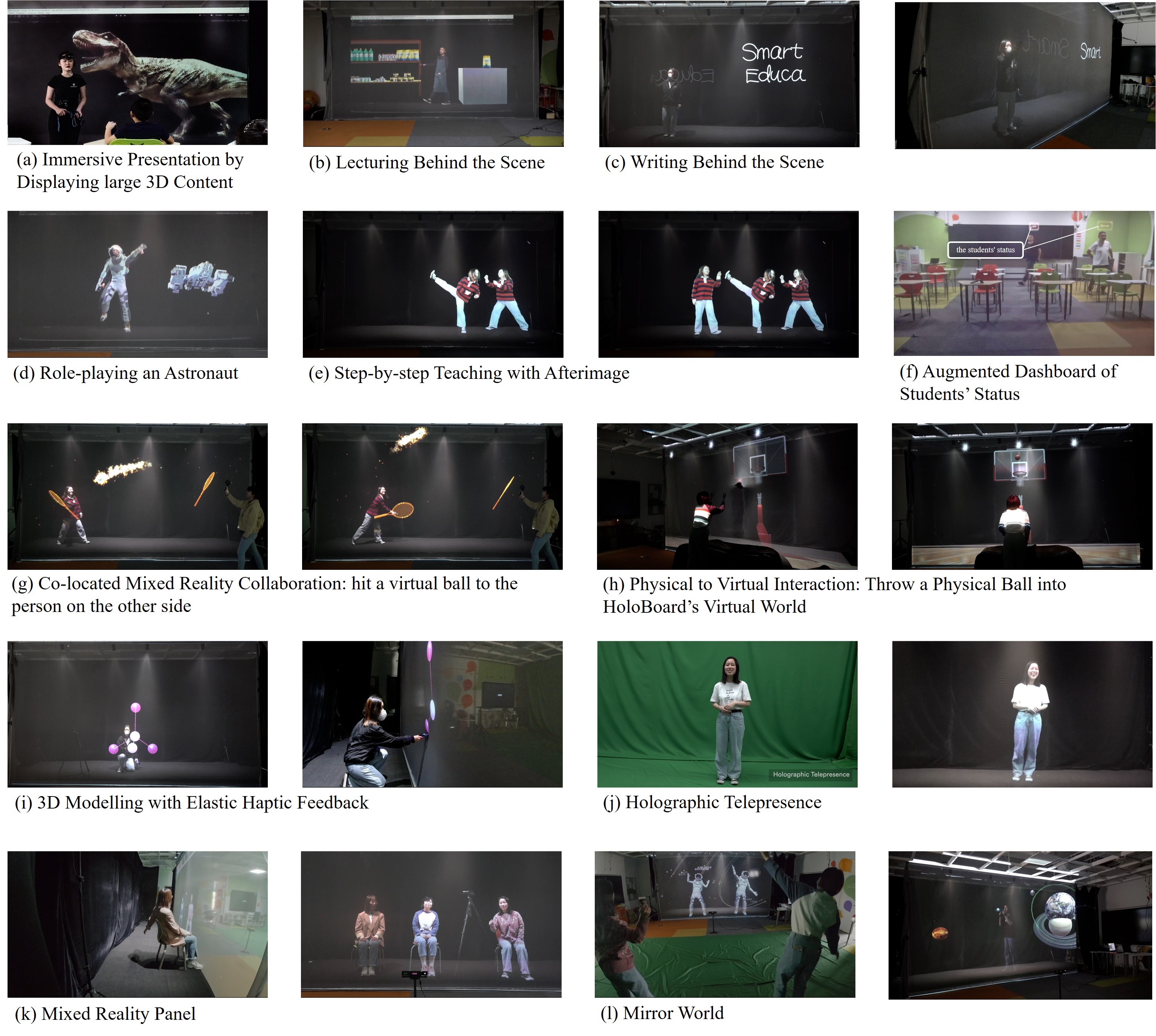}
  \caption{Implemented Scenarios of HoloBoard: (a) Immersive Presentation by Displaying large 3D Content; (b) Lecturing Behind the Scene; (c) Writing Behind the Scene; (d) Role-playing an Astronaut; (e) Step-by-step Teaching with Afterimage; (f) Augmented Dashboard of Students’ Status; (g) Co-located Mixed Reality Collaboration: hitting a virtual ball towards the person on the other side; (h) Physical to Virtual Interaction: throwing a physical ball into a virtual world displayed by HoloBoard; (i) 3D Modelling with Elastic Haptic Feedback; (j) Holographic Telepresence; (k) Mixed Reality Panel; (l) Mirror World.}
  \label{keyfeatures}
\end{figure*}

\subsubsection{Double-sided Rendering for HoloBoard}

The key to creating an immersive holographics lies in real-time rendering of the position, size and angle of the content with reference to changes in the observer's position and perspective. We achieved this by converting the spatial position into the camera coordinate system of Unity, and calibrating the orientation and scale of the digital contents to be rendered based on the user's position. Experimentally a point in the center of the classroom was selected as the "best" viewing spot for the students and teacher. It simplified the calculations while guarantying the rendering respond to the user's position and movement accordingly.

\subsubsection{Implemented Scenario}
According to the interaction technique design in the former section, we implemented 12 scenarios for the working prototype as showing in Fig.~\ref{keyfeatures} and our demo video.

\begin{enumerate}[(a)]
 \item Immersive Presentation by Displaying large 3D Content: When the teacher clicks the next/previous page on the controller, it will automatically jump to the next/previous slide.
 \item Lecturing Behind the Scene: The teacher can manipulate the digital content flowing between the teacher and students by one or both controllers, with eye contact with students at the same time.
 \item Writing Behind the Scene: The teacher can write freely with one controller behind the scene. The system displays the original text using a inconspicuous color for the presenter and generates another text reversal to display the correct text using a conspicuous color for the audience simultaneously.
 \item Role-playing an Astronaut: The spatial positions of the controllers and trackers worn by the teacher are mapped to a 2D/3D avatar model with an animated skeleton structure. When the teacher moves, the 2D/3D avatar model will be rendered on the projected screen right in front of the teacher and copy the teacher’s movement. This function is accomplished by converting spatial coordinates into project coordinates based on students’ point of view. From the students’ view, the teacher is augmented by the avatar outfit, which can also be used to trigger more animations.
 \item Step-by-step Teaching with Afterimage: The teacher can make a pose and click the button on the controller to freeze his/her image/avatar on the current page of HoloBoard.
 \item Augmented Dashboard of Students’ Status; By tracking the position of the teacher's eye and students, each student’s name tag and other real-time learning information will appear above each student’s head. Furthermore, the teacher can give marks (like praise) to a certain student by clicking on the student’s name tag. 
 \item Co-located Mixed Reality Collaboration: The user can pass a virtual object towards another user on the other side by the controller with naturally eye contact at the same time, such as playing tennis, collaborative manipulation;
 \item Physical to Virtual Interaction: The students can throw a physical ball into a virtual world displayed by HoloBoard. We tied a vive tracker to the ball for tracking.  When the physical ball contact the screen, the system create
 \item 3D Modelling with Elastic Haptic Feedback: The user can create a new model in Z-direction on the screen by hold the controller into the HoloBoard.
 \item Holographic Telepresence: The remote user can be photographed by the camera in real-time in front of the green screen, and after removing the background, it will be displayed vividly on the HoloBoard. 
 \item Mixed Reality Panel: The local users can sit behind the HoloBoard and conduct a mixed reality panel, discussion with remote users' vivid telepresence on the HoloBoard. 
 \item Mirror World: The users can be photographed/captured by a camera, and blended, re-positioned, and resized into the digital mirror scene displayed by HoloBoard.
 \end{enumerate}

\section{Experimental Study}

\begin{figure}[h]
  \centering
  \includegraphics[width=\linewidth]{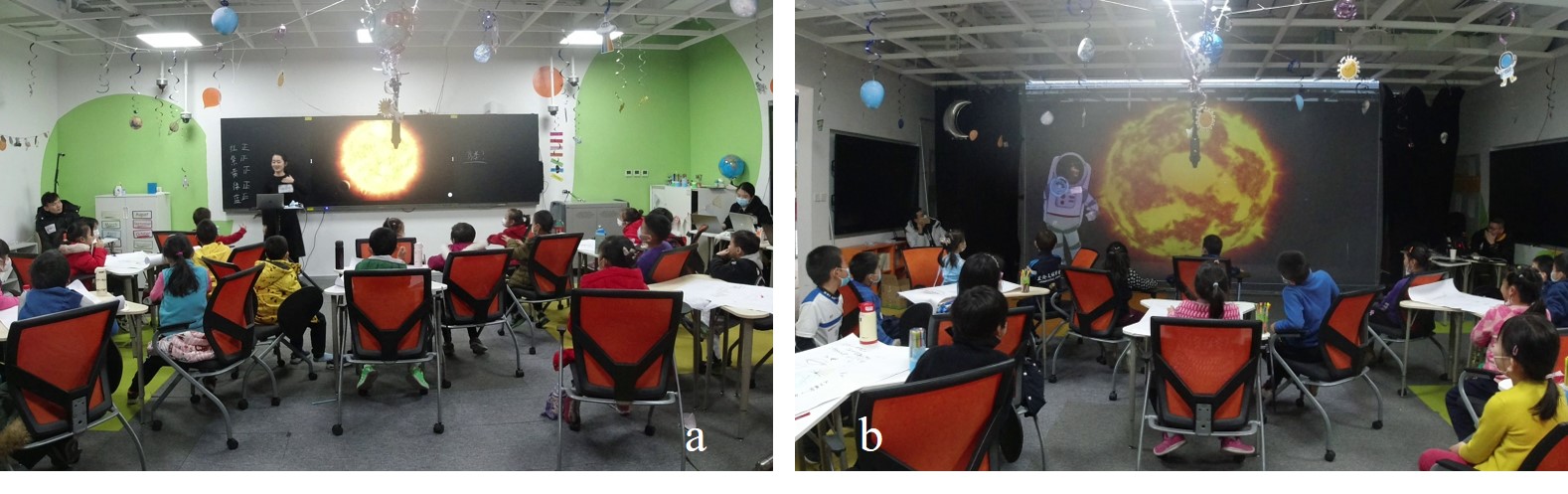}
  \caption{Compared Experiment Environment set up: a) Normal group; b) HoloBoard group}
  \label{experiment}
\end{figure}

To investigate whether HoloBoard can actually facilitate children's learning (e.g., engagement and learning outcome) and explore the additional value it can offer to a lecture-based class, a controlled experimental study was conducted.

\subsection{Course Design}
HoloBoard is designed as an essential teaching board, while having a form different from the normal interactive whiteboards. We conducted a participatory course design process inspired by \cite{penuel2007designing} to amplify the advantages of HoloBoard and design a course for experimental study.

Five professional teachers (all females, avg. teaching experiences = 4.2 years in Literature, STEM or Arts in primary school, middle school or high school) got invited. HoloBoard's designers intensively conducted in-person and online participatory design workshops (over four formal workshops, each lasting 1-2 hours) within 6 months (Jul 2020 - Dec 2020). Moreover, we collected the syllabuses, courseware, video recordings of lectures, and student work from previous similar courses from these participating teachers to better understand user scenarios. 


After rounds of discussion, teachers and researchers reached an agreement that topics about the outer space can be highly interesting to young students. According to Piaget's theory of cognitive development \cite{huitt2003piaget}, children aged 6-9 are at the Concrete Operational Stage. In this stage, in general, children's abstract thinking skills are still developing and they can only apply abstract concepts, (e.g., time, space) in concrete real-life situations. Therefore, with the large-format and highly interactivity, HoloBoard has the potential for facilitating children's abstract learning by displaying the course content (i.e, space) in an intuitive way, benefiting from 3D visual effects of planetary, interactive experiment platform, and situated role-play, etc.  

Eventually, a STEM course (i.e. Space Exploration) targeting six- to nine-year-old students was designed. The course focused on the characteristics of the eight planets in the solar system. In the course, students were invited to explore the eight planets in the solar system in order to find a habitable planet. Students were guided to think about questions including "why should humans explore Mars?" and “what kinds of planets are habitable?" The course adopted the inquiry-based science teaching method and was designed according to the 5E Instructional Model process \cite{duran20045e}. 

Three main features of HoloBoard were adopted into the course: 1) the teacher can role-play as an astronaut to roam the solar system, and guide the students to learn the characteristics of the eight planets situationally; 2) the teacher can double-sided demonstrate the order of the eight planets interactively using virtual experiment simulation; 3) the students can double-sided play multi-user game to test their knowledge about the characteristics of the eight planets. In the normal group, to maintain the fairness of the comparison, the script of the classes was kept the same and the following adjustments to activities were made. For Feature 1, (AR-based role-pay), in the normal group, the teacher also verbally pretended to be an astronaut (asking the students to imagine that she was an astronaut) and thus kept the teaching script the same. For Feature 2 (double-sided demonstration), during learning the order of the planets, a similar demonstration of planet order was displayed on the interactive whiteboard in the control class (as depicted in Figure \ref{videoSample}(b)(e)). For Feature 3 (double-sided multi-use game), while summarizing planetary habitability, the teacher mainly posed guiding questions (e.g., Is Mars a good place to live in?) and asked students to analyze from different dimensions (e.g., gravity and dangerous factors) in the two classes. The only difference is that in the normal group, a table summarizing the planet habitability was used while the interface of the VR multi-user serious game was used in the HoloBoard group.

\subsection{Site Preparation}

Besides HoloBoard setup, a traditional classroom with a regular interactive whiteboard (4K, 86-inch) was set up. Desks and chairs were arranged in five groups. 
To facilitate data gathering, six cameras were set up around the classroom. One 8K panoramic camera (\textit{TECHE360anywhere}) was set up in the middle of the classroom. An audio recording device (\textit{iFLYTEK SR501}) was also set in the experimental classroom. 
Furthermore, this research received IRB approval from our local institution.

\subsection{Participants}

36 participants (Female = 18, 15 in Grade 1, 15 in Grade 2, and 6 in Grade 3, avg. age = 7.06) were selected upon a pre-experiment survey and a pre-test knowledge test. Most participants reported being exposed to digital devices like projectors, mobile phones, tablets, and PCs before. Approximately 78\% of the participants mentioned preferences for these smart digital devices. Moreover, the participants were fairly interested (4.3 out of 5) in space related topics. Their motivation, engagement, and learning self-efficacy on average were at a slightly high level (15.5 out of 20).

Besides, we invited one of the teacher subjects in the course design session to be our experiment teacher (T7, STEM teacher, 26 years old, female, master's degree in Education, 2 years of teaching experience). This teacher teaching both HoloBoard group and normal group with same class script.

\subsection{Procedure} 



We pairwise assigned the participant students (based on their gender, age, grade, performance on the pre-test, interests in related topics, learning attitude, and past experience with technology) to either a control group (normal group) subjected to the traditional practice based on the normal lecture-based class, or an experimental group (HoloBoard group) subjected to the practice within HoloBoard environment. In addition, we discussed the course outline with the teachers to further control their lecturing script for the two classes. 


Before the experiment, parents/legal guardians of the participants all signed the informed consent release form. During the experiment, the two groups of students were taught the same 70-minute lecture featuring either the HoloBoard or the normal interactive whiteboard. The participants were seated in assigned small groups. After the experimental class ended, the experimenters conducted a one-on-one post knowledge test with each student participant. Each post-test took approximately 20 minutes.





A pre-/post- knowledge test (See Appendix) was designed, identical for all participants, to check students' knowledge of class content before and after the lecture-based class. The test consisted of 12 questions of different complexity and types (four single-choice questions, two multiple-choice questions, and six open-ended questions), each of which covered one or more cognitive learning objectives of the class. The test was developed following the examples of usual primary school tests and was validated by the teachers. Each single-choice question is worth one point and each multiple-choice question is worth two points. The score for open-ended questions can range from 0 to 5 based on the rating rubric used in \cite{lui2018designing}. Thus, the full score of the pre-/post knowledge test was 38.


\subsection{Data Collection and Analysis}
Our data came from three primary sources: 1) the pre- and post-knowledge tests, 2) audio recording and transcripts of the HoloBoard class and the normal class, and 3) video recordings of the two classes. The knowledge test was used for learning outcome comparison analysis and the other two were for student in-class engagement comparison analysis.

\subsubsection{Pre-/post- Knowledge Test Data.}
Students' answers to the single/multiple choices questions in the pre- and post- knowledge tests were scored by two researchers based on the right answers provided by the teacher. In addition, students' answers to the open-ended question were audio recorded, transcribed, and rated by two trained coders based on a rating rubric adapted from the Knowledge Integration rubric used in \cite{lui2018designing}. Two trained coders independently coded all students' responses. The inter-rater Kappa was greater than 0.70 (p < .001). A third coder reviewed their ratings, and consensus were eventually achieved. Then the total scores of each each students' pre- and post-tests were calculated and compared.

\subsubsection{In-class Engagement Learning Data and Analytics.} The learning data included video and audio recordings and transcripts of the two classes. 

\textbf{Video Data and Analysis.}
The video recordings of the two classes were mainly used for student behavioral and engagement analysis by manual coding.
The coding scheme was adapted and integrated from \cite{becsevli2019mar,homer2018comparing,hsieh2015exploring,pas2015profiles,wagner2005physiological}, which pictured students’ reactive behavior to the two teaching devices and reflected students' engagement in the classes. The three main categories of our codes are affective states (emotional engagement), classroom behavior and posture (behavior engagement). The detailed types and descriptions, and operational definition of students' nonverbal behaviors are presented in Table \ref{tab:NonverbalCode}.

\begin{figure*}[h]
  \centering
  \includegraphics[width=  \linewidth]{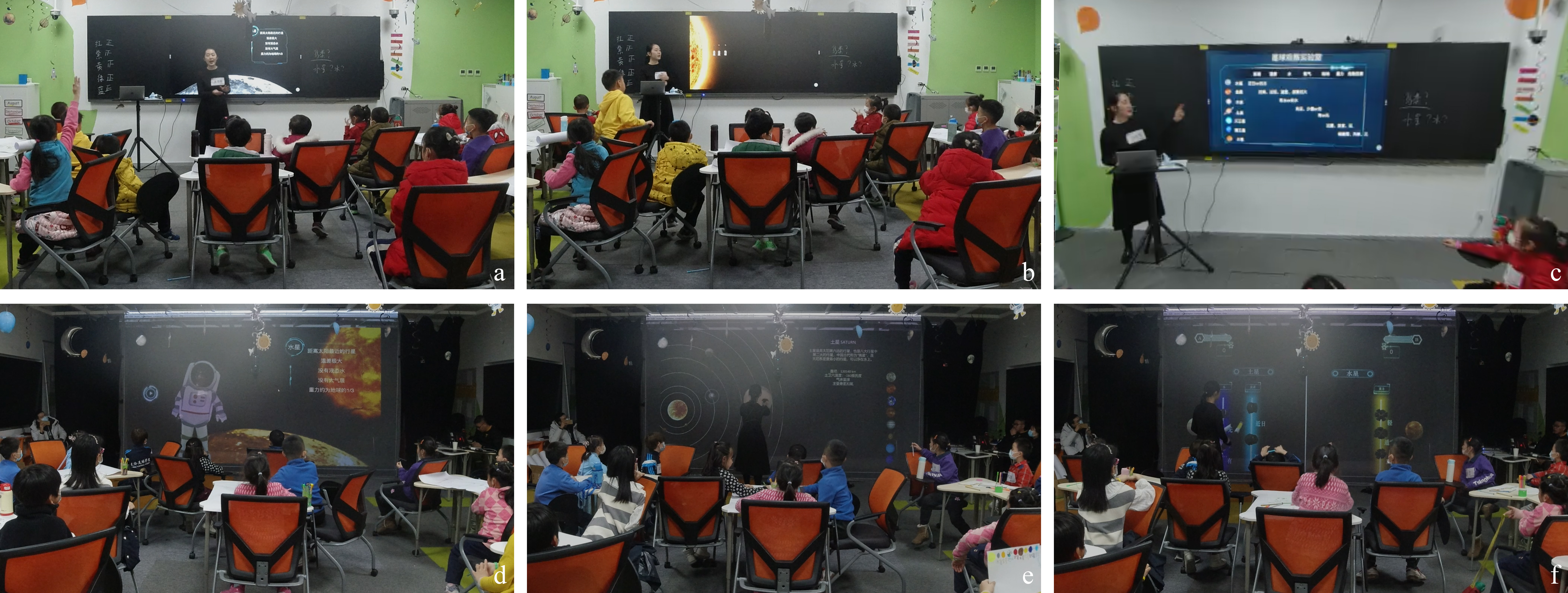}
  \caption{Six Video Samples of Three Comparable parts in two Classes: Exploring planets’ characteristics(a)(d); Discussing the order of planets(b)(e); Summarizing planetary habitability(c)(f)}
  \label{videoSample}
\end{figure*}

We selected three video clips from the HoloBoard and the normal classes, respectively, discussing the same topics (exploring planets’ characteristics, discussing the order of planets, and summarizing planetary habitability). While discussing the first two topics, in the HoloBoard group, the two key features of the HoloBoard---AR-based role-play and VR simulator demonstration---were used respectively. As noted in the course design section, these activities remained comparable because they used similar scripts and both involve elements of role play and made use of the same images.
To keep the comparability between the two classes, the video clip of children playing the multi-user serious game on HoloBoard was excluded from the analysis.
Furthermore, three random 10-second video clips were sampled and analyzed from the two classrooms’ six video recordings (see fig.\ref{videoSample}). The random sampling of short video clips has been widely used in prior literature for analyzing student interaction (e.g., \cite{ groccia2018student} and \cite{karsenti2016interactive}). Each participant was analyzed through 90-second video data in which their nonverbal behavior was coded. All video clips were analyzed using the Noldus Observer XT software. While coding the video clips, the type of nonverbal behavior of each student mentioned in Table \ref{tab:NonverbalCode} was identified and marked by the coder in the Observer software, and the duration of each behavior was automatically calculated and computed by the software. To calculate the length of each behavior, our unit of time was defined as 0.1 seconds. Two trained coders coded each student’s behavior based on the coding scheme independently. Inter-rater reliability was conducted on 22\% of the video data, where the inter-rater Kappa was greater than 0.89 (p < .001).

\begin{table*}[]
\small
\caption{Manual Coding Scheme of Video Recordings}
\begin{tabular}{p{50pt}p{100pt}p{230pt}}
\hline
Categories   & Types & Description   \\\hline
\multirow{4}{*}{Affective States}
    & High Arousal Positive Valence   & \shortstack{Joy: happy, excited, aroused, concentrated, etc.}   \\
    & Low Arousal Positive Valence     & \shortstack{Calm: relaxed, satisfied, and content, etc.} \\ 
    & High Arousal Negative Valence    & \shortstack{Anger: tense, annoyed, frustrated, etc. }  \\
    & Low Arousal Negative Valence   & \shortstack{Boredom: bored, depressed, miserable, etc. }\\\hline
\multirow{3}{*}{Posture}
    & Close Posture   & \shortstack{Moving closer to or leaning toward the whiteboard/HoloBoard}\\
    & Neutral Posture    & Sitting straight up\\
    & Leave  Posture  & Moving body away or leaning away from the whiteboard/HoloBoard\\\hline
\multirow{3}{*}{Behavior} 
    & Positive Behavior   & Engaged behavior: answering questions, raising hands, looking at the pupil speaking, applauding, etc.\\
    & Normal Behavior   & Normal behavior: listening, looking at the screen/teacher, writing, etc.\\
    & Misbehavior  & Disengaged behavior: chatting, looking away etc.    \\
    \hline                                                                                
\label{tab:NonverbalCode}
\end{tabular}
\end{table*}

\textbf{Audio Data and Analysis.}
The HoloBoard and the normal groups’ classroom discourses were recorded, fully transcribed, and analyzed. The audio recordings were used to conduct an acoustic analysis of frequency and loudness as references to student emotional engagement. The acoustic analysis was performed in pyAudioAnalysis. To better portray the change of student engagement and compare the results of the two groups, we selected and computed the first quarter of data of the two classes as a baseline. 

\textbf{Transcript Analysis.}
The audio recordings of the two classes were fully transcribed. The transcripts were firstly used to compare the teachers' speech type in order to confirm the effectiveness of controlling teachers' lecturing script. The coding scheme of teachers' speech was mainly adapted from FIAS \cite{flanders1970analyzing}. As shown in Table \ref{tab:VerbalCode}, teachers' initiated language was categorized into three main types: lecturing (e.g., giving information), directing (giving direction or instructions), and asking questions. Then, questions were further divided into close- or open-ended questions as previous research has shown that teachers' questions have influence on student cognitive engagement \cite{smart2013interactions,Murcia2010TalkingAS,lee2012teacher}. 

Focusing on students, we used the transcripts to analyze cognitive and emotional engagement by both manual coding and algorithm processing. For cognitive engagement analysis, firstly, students’ speech types (i.e., statement and questions) and cognitive levels were coded. For student speech type, the coding scheme was adapted from the linguistic expression category of the scheme in \cite{mulder2002assessing}. For the cognitive level analysis, Bloom’s taxonomy of cognitive domain \cite{bloom1956taxonomy} was used and thus the cognitive level of students’ responses were categorized into: remembering, understanding, applying, analyzing, evaluation, and creation. Four trained coders coded the transcript of the two classes, and the inter-coder Kappa was greater than 0.81 (p < .001).

Moreover, the whole transcripts were used to compare the attributes of discourses as references to cognitive engagement. The discourse attributes we investigated include: (a) the number of speaking turns, (b) the number of times each student spoke, (c) the number of words and sentences each student spoke, and (d) the length of students' sentences and rounds of dialogue, which were adapted from the related studies (e.g., \cite{stroud2015learner,serban2017interactive,black2009comparison}).

To measure emotional engagement, we captured students' exclamations (e.g., “Wow” and “Oh My God”), which was captured to assess students' engagement in \cite{becsevli2019mar}. In addition, we also used Baidu AI Cloud \cite{tian2020skep} to perform semantic analysis as indicators of emotional engagement.

\begin{table*}[]
\small
\caption{Manual Coding Scheme of Transcripts}
\begin{tabular}{llll}
\hline
Speaker & Categories & Types & Description \\\hline
\multirow{4}{*}{Teacher} & \multirow{4}{*}{Speech Type}
& Lecturing & Giving information or opinion\\
&& Directing & Giving directions\\
&& Asking Close-ended Question & Calling for a single response\\
&& Asking Open-ended Question & Calling for board responses\\\hline
\multirow{10}{*}{Student}
&\multirow{3}{*}{Speech   Type}
    & Assertion             & Statements of facts, opinion, choices, etc.   \\
    && Question              & Request for information   \\
    && Exclamation           & Sound/Words expressing strong emotion (e.g., “Wow” and “Pff”)   \\\cline{2-4}
&\multirow{6}{*}{Cognitive Level}
    & Remembering           & Reciting and memorizing labels    \\\
    && Understanding         & Relating and organizing previous knowledge    \\
    && Applying              & Applying information into a new situation   \\
    && Analyzing             & Drawing connections among ideas   \\
    && Evaluation            & Examining information and make judgement    \\
    && Creation              & Creating new ideas using what has just been learned   \\\hline
\label{tab:VerbalCode}
\end{tabular}
\end{table*}

\subsection{Results}
We firstly compared the teacher’s script in the HoloBoard and normal classes. Teachers' questioning accounted for approximately six percent of the speech, among which 70\% of the questions were close-ended questions in both classes. The results suggested there were no major differences in the types of speech initiated by the teacher between the two classes. Then we describe the results of students' learning outcomes and engagement as the following.

\subsubsection{Learning Outcomes. }
 
We conducted Mann-Whitney U test to compare the knowledge pre-test results as \emph{baseline} between the normal and HoloBoard groups. The result suggested there was no significant differences between the two groups ( p = 0.46).

Next we compared learning gains and performances between the two groups. The analysis results of learning outcomes are presented in Table \ref{tab:testscore}. For the learning gains, the post-test score of the HoloBoard group (Mdn = 20.5) was significantly higher than their pre-test score (Mdn = 26.5), U = 56.5, p < .001. In addition, the effect size (r = 0.78) was calculated as a rank-biserial correlation coefficient, suggesting the difference is large (r = 0.78). For the normal group, the post-test score (Mdn = 24) and pre-test score (Mdn= 21) also had significant difference (U = 105, p = .04). A medium effext size (r = 0.42) was found. 

As for comparing the post-test scores of the two groups, the descriptive results suggested the HoloBoard group achieved a higher mean score (M = 26.39) in the post-test than the normal group (M = 24.61), although their mean pre-tests scores were rather similar (20.94 and 20.79, respectively). However, independent samples Mann-Whitney U test results indicated that the difference was not statically significant, U = 125.5, p = .13.

\begin{table}[]
\small
\caption{Pre-/post-test Results (mean and standardized deviation) of Normal and HoloBoard group.}
\begin{tabular}{lllll}
\hline
    & HoloBoard (n=18) &  Normal (n=18) & Effect size       \\\hline
Pre-test & 20.10 (3.85)   & 20.78 (6.08 )                              & 0.02 \\
Post-test  & 26.40 (4.59)    & 24.61 (5.58)                           & 0.19 \\
Effect size  & 0.78 ***     & 0.42*                           &   \\\hline
\label{tab:testscore}
\end{tabular}
\end{table}

\subsubsection{Student In-Class Engagement.} 

Students in HoloBoard group showed higher levels engagement in all the three dimensions: behavioral, emotional, and cognitive engagement according to our multimodal learning analytic results, which are shown in
Table \ref{tab:multimodalResults}.



\begin{table*}[]
\caption{Results of Student Engagement Analysis}
\begin{tabular}{p{35pt}p{130pt}p{70pt}p{60pt}p{40pt}p{30pt}}
\hline
Data   & Variable  & HoloBoard Group & Normal Group & Value & \emph{p}\\\hline
\multirow{10}{*}{Video}
       & Close Posture & \textit{Mdn = 40.5**} & \textit{Mdn = 11.0} & \textit{U = 257.5}  & 0.002\\
       & Neutral Posture & \textit{M = 38.28}  & \textit{M = 55.78*} & \textit{t = 2.67}   & 0.013\\
       & Leave Posture   & \textit{Mdn = 8.0}  & \textit{Mdn = 14}   & \textit{U = 156}    & 0.864\\ 
       \cline{2-6}
       & Positive Behavior& \textit{Mdn = 3.5*}& \textit{Mdn = 0}    & \textit{U = 235.5}  & 0.019\\
       & Normal Behavior & \textit{Mdn = 82.0} & \textit{Mdn = 77.5} & \textit{U = 183.5}  & 0.501\\
       & Misbehavior & \textit{Mdn = 8.0} & \textit{Mdn = 11.5} & \textit{U = 125.5}  & 0.252\\
       \cline{2-6}
       & High Arousal/Positive Emotion & \textit{Mdn = 4.0*}& \textit{Mdn = 0}& \textit{U = 234.5}& 0.02\\
       & High Arousal/Negative Emotion & \textit{Mdn = 0}   & \textit{Mdn = 0}& \textit{U = 124.0}& 1\\
       & Low Arousal/Positive Emotion  & \textit{Mdn = 85.5}& \textit{Mdn = 79.0} & \textit{U = 193.0} & 0.339\\
       & Low Arousal/Negative Emotion  & \textit{Mdn = 0}   & \textit{Mdn = 11.0*}& \textit{U = 82.5}& 0.011\\\hline
\multirow{2}{*}{Audio}
       & Loudness & \textit{M = 0.000231***} & \textit{M = 0.000153} & \textit{t = 7.64} & \textless .001\\
       & Frequency& \textit{M = 0.000032}    & \textit{M = 0.00017}  & \textit{t = -0.60}& 0.27\\\hline
\multirow{15}{*}{Transcripts}
       & Speaking Turns/min  & \textit{8.14}& \textit{7.23}& \textit{/} & \textit{/} \\
       & Mean Sentence Length & \textit{M = 2.29}& \textit{M = 2.35}& \textit{t = 0.564}& 0.28\\
       & Mean Speaking Turn Length & \textit{M = 2.42}& \textit{M = 2.54}& \textit{t = 0.840} & 0.2\\
       & Student Speaking Times/min & \textit{M = 3.24***} & \textit{M = 1.59}& \textit{t = 26.155}&\textless .001 \\
       & Words Spoken per Student/min & \textit{M = 7.86***} & \textit{M = 4.10}& \textit{t = -9.456}& \textless .001\\
       & Sentences Spoken per Student/min & \textit{M = 3.43***} & \textit{M = 1.72} & \textit{t = 20.892} & \textless .001\\\cline{2-6}
       & Assertion& \textit{Mdn = 3.05***}& \textit{Mdn = 1.6}& \textit{U = 308.0}& \textless .001\\
       & Questions & \textit{Mdn = 0.027**}& \textit{Mdn = 0}& \textit{U = 250.0}& 0.005\\
       & Exclamations & \textit{Mdn = 0.08***} & \textit{Mdn = 0.07} & \textit{U = 324.0} & \textless .001\\\cline{2-6}
       & Remembering & \textit{Mdn = 1.54***} & \textit{Mdn = 0.53} & \textit{U = 316.0} & \textless .001\\
       & Understanding & \textit{Mdn = 0.46***} & \textit{Mdn = 0.32} & \textit{U = 279.0} & \textless .001 \\
       & Applying & \textit{Mdn = 0} & \textit{Mdn = 0} & \textit{U = 124.0} & 0.239 \\
       & Analyzing & \textit{Mdn = 0.08} & \textit{Mdn = 0.12} & \textit{U = 116.0}  & 0.152\\
       & Evaluation & \textit{Mdn = 0} & \textit{Mdn = 0} & \textit{U = 144.0} & 0.584 \\
       & Creation & \textit{Mdn = 0} & \textit{Mdn = 0} & \textit{U = 170.0} & 0.815\\\cline{2-6}
       & Positive Emotion& \textit{M = 1.36***} & \textit{M = 1.20} & \textit{t = -14.22} & \textless .001\\
       \hline
       \label{tab:multimodalResults}
\end{tabular}
\end{table*}

\textbf{Behavioral Engagement Results}
Generally, students in the HoloBoard group were more engaged behaviorally than those in the normal group, indicated by more close posture and more positive classroom behavior.
Mann-Whitney test indicated students in the HoloBoard group demonstrated significantly longer duration of close posture (p = .002) and positive behavior (p = .019) in the HoloBoard group than students in the normal group. 
 
\textbf{Emotional Engagement Results}
Students in the HoloBoard group exhibited higher levels of emotional engagement, which was demonstrated by more emotion that has high arousal and positive valance. This result was based on video and audio data coding, acoustic analysis, and semantic analysis.

For video coding results, the length of high arousal and positive emotion was significantly larger in the HoloBoard group (p = .02). We also found a significantly smaller length of low arousal and negative emotion in the HoloBoard group (p = .01). The audio data coding results also showed that students were more emotionally engaged in the HoloBoard class: the number of student exclamations per minute in the HoloBoard class was significantly larger than the one in the normal group (p < .001). Moreover, t-tests were conducted to compare voice loudness and frequency of the HoloBoard and normal groups. There was a significant higher level of loudness in the HoloBoard group (p < .001), whereas there was no significant difference in frequency between the two groups, indicated by the zero-crossing rate (p = .27). The greater voice loudness indicated higher levels of arousal in emotion of students in the HoloBoard group.

In addition, t-test was performed on the semantic data of the two groups' transcripts. The results indicated there were significant more positive emotion (p < .001) and  probability of positivity (p < .001) in the HoloBoard group than in the normal group. These results further confirmed the manual coding and acoustic analysis results, suggesting that the emotional engagement level of students in the HoloBoard group was higher. 

\textbf{Cognitive Engagement Results}
In general, students in the HoloBoard group showed more signs of cognitive engagement compared to those in the normal group. The signs included more contribution to classroom discussion as well as more responses in the remembering and understanding levels. Firstly, Mann-Whitney U tests were conducted on types of students’ speech. The results indicated that the numbers of students’ assertions (p < .001) and questions (p = .005) per minute were both significantly larger in the HoloBoard group than the one in the normal group. 
 
In addition, t-test results of comparing discourse attributes suggested the number of student speaking times in the HoloBoard group was significantly greater than the one in the normal group (p < .001). In addition, the numbers of words and sentences students spoke per minute in HoloBoard group were also significantly larger than those in the normal group (both p-values are smaller than 0.001). There was no significant differences in other attributes. The results further confirmed the speech type analysis results, suggesting students in the HoloBoard group were more cognitively engaged and more willing to contribute to class discussion.
 

For the cognitive level of students' responses, the numbers of remembering responses (p < .001) and understanding responses (p < .001) were significantly higher in the HoloBoard group than in the normal group. No significant differences were found in the number of responses at any higher cognitive level (e.g., applying and analyzing responses) between the two groups.

\section{Discussion}
\subsection{Engaging Students with HoloBoard}

The results of the experimental study show that HoloBoard can lead to significantly higher levels of engagement (i.e., behavioral, emotional, and cognitive) and slightly better learning outcomes in comparison to the normal class. 
Specifically, throughout the class, students in the HoloBoard group were found more active to participate in the class. The results indicated that students demonstrated more concentrating and positive behavior, higher levels of valance and arousal of emotion, more intensive rounds of discussion, and more remembering and understanding responses in the HoloBoard group than in the normal group. 
These signs of behavioral, emotional, and cognitive engagement observed in our study were consistent with the ones identified in previous literature, e.g., \cite{becsevli2019mar,hsieh2015exploring,fredricks2004school}.

In particular, the techniques of the HoloBoard elicited more engaging activities from the students during the discussions. For instance, while studying the characteristics of planets with the role-play in the HoloBoard class, students consistently came up with ideas for the teacher to explore the space as an astronaut. 
Some of them, such as "try to jump (on Mars)," and "try to weigh yourself (on Mercury)" indicated the students' intentions to apply their new knowledge of space (e.g., change of gravity) and their collective efforts to make meaning out of the information~\cite{smart2007learning,srivastava2014role,davidson2014boundary,matthews1996collaborative}, which is the sign of students' active learning \cite{smart2007learning,srivastava2014role} and collaborative learning \cite{davidson2014boundary,matthews1996collaborative}.
While using the virtual simulator demonstration to learn the order of planets, students were willing to share their opinions and get engaged in debates (e.g., where should the Earth be). 
They also exclaimed with excitement (e.g., "Wow" and "Yeah! We did it!") when all the planets were placed in order. 
These anecdotes demonstrated that the usage of the HoloBoard created an engaging class where students could readily and actively participate and contribute. 

In conclusion, with HoloBoard, the lecture-based class managed to create a more engaging atmosphere for active student participation.

\subsection{Design Implications}
\subsubsection{Challenge of Role-play in Lecture-based Classes.}
Although the augmented role-play function was welcomed by both teachers and students, there have raised other concerns. With our observation, proportion of teacher's close-ended questions dramatically increased while using role-play, which caused the number of the students' lower cognitive level in responses to increase. After simplifying the enter/exit action for role-play (i.e., we set a shortcut key link to this function from the original secondary menu.), the proportion of the experiment teacher's close-ended questions returned to normal in the experiment study, and the experiment teacher's self-report also confirmed the effectiveness of this change. 

This suggested potential challenges when involving novel techniques such as role-play in lectures. 
On the one hand, unfamiliar technology increased teachers' extraneous cognitive load and affected their teaching presence, as highlighted by previous research \cite{kerawalla2006making}. Therefore, the ease of use and necessary training are important factors in adopting novel lecturing techqniues
On the other hand, role-play, as an effective teaching method, could be demanding for an instructor, both in preparation and in implementation \cite{ardriyati2009roleplay}. Hence, when using role-play functions with HoloBoard in class, the system needs to support teachers (i.e., using AI assistants to help classroom management and learning process management) and allow them to control the pace of role-play easily. 

\subsubsection{Simulator Demonstration and Multi-user Game in Classroom.}
We found that the simulator demonstration and the game brought particular excitement to the classroom. In the classroom, a very simple simulation demonstration of the planetary order not only attracted the attention of the students but also stimulated discussion on additional knowledge beyond the syllabus, for example, the orbital speed of the planets. Indeed, the experimenter teacher confirmed such positive effects and recalled that "everyone was so excited when I finally put all the planets in the right order." She was impressed by the visual and intuitive way of displaying content materials and believed that "such direct observation and perception of orders and orbital speed of the planets can have amazing effects."

In addition, the multi-user game, enabled by the large-format screen of the HoloBoard, also resulted in more learning opportunities for students. In the post-test self-reports, the students in the HoloBoard group had a significantly higher collaborative-learning-related self-evaluation score. They also expressed that they felt their teams had more trust in other members, and they had more frequently encouraged other members to join the group activities. The experimenter teacher also thought "the game allowed students to apply the knowledge they just learned." 

In addition, we agree that the interactive designs of HoloBoard should be able to sustain student engagement in the long run. Although our current study only compared immediate influence, we believe with the simulator demonstration of content knowledge and gamification elements discussed above, have the potential to overcome the novelty effect. because we have embedded pedagogical factors in our design, for example, selecting the appropriate knowledge to stimulate and presenting the learning journey mading up of raising questions, exploration, and end game. Such ways to make the engagement meaningful and helpful for students are suggested ways to overcome the novelty effects in \cite{tsay2020overcoming}.

In summary, simulator demonstrations and games enabled with HoloBoard can lead to positive student experiences in multi-user learning environments and we recommend including more such activities in classrooms.

\subsubsection{Interactive Transparent Screen for Children}

We offered a free-play session where students could play the virtual multi-user game on the HoloBoard after the class. We observed that the students made good use of the double-sided feature of the HoloBoard and engaged in cooperative pretend play on the two opposite sides of the HoloBoard spontaneously. For example, in one of the pretend play scenarios, some students in front of the HoloBoard waved controllers in the air and create the visual and audio effects reflecting the hitting of stones, pretending there were attacking students behind the HoloBoard. Those students behind the HoloBoard traced the location of the effects and pretended that they got hurt and fell down.  

Pretend play has long been recognized as a vital type of children's play from the perspective of children's cognitive, social, and physical development \cite{vygotsky1980mind,lillard1993pretend}. Furthermore, children who engaged in more pretend play generally have more advanced social-cognitive skills, including social perspective taking \cite{fein1981pretend} and social competence \cite{connolly1984relation}. Hence, the development of children' higher-level cognition is facilitated \cite{bergen2002role}.

HoloBoard showed strong potential as a platform for active and beneficial interaction between children. Beyond classrooms, such interactive transparent displays could be installed at children's playgrounds to facilitate more pretend play and social play via the double-sided interaction feature.

\subsection{Limitation and Future work}

For the HoloBoard system, regarding the expectation of potential implementation in the classroom settings, there are several concerns such as a dimmer display color compared to ordinary LED displays, the requirement of trackers and controllers, and so on. Nevertheless, most participants expressed their expectations to see HoloBoard in actual classrooms in the near future. In addition, since the hardware cost and maintenance of HoloBoard is manageable, we will focus on optimizing creative tools like Chalktalk \cite{perlin2018chalktalk} and body-driven augmented graphics creative tool \cite{saquib2019interactive} to make it feasible for teachers to independently develop content like slides or role-playing in the future.

For the study, the current experimental study did not fully utilized the functions of HoloBoard. This was for several reasons. First, it was non-trivial to design a course that includes all the novel functions of the system. Interaction techniques such as "augmented dashboard", "physical to virtual interactions", "3D modeling", and "holographic telepresence" shall be designed within other course contents and lecturing contexts. Second, it is fairly time-consuming for the experimental teachers to understand and get used to the new functions. Therefore, how to best integrate these techniques of the HoloBoard systems into a lecture requires further explorations. Moreover, we acknowledge that the current study design comparing student interaction during the two classes can not help us estimate possible novelty effects. Verifying and reducing possible novelty effects in long-term usage will be our future work.

Meanwhile, due to individual differences among learners and novelty effects of new technology, it remains a challenge to exploit the advantages of such immersive technology in wild. In the future, we plan to leverage HoloBoard in long-term use cases and expand on a larger scale.

\section{Conclusions}
We presented HoloBoard, a large-format immersive teaching board based on pseudo holographics. Its unique features of immersive visual presentation and transparent screen allowed us to explore a rich set of novel interaction techniques, potentially valuable for lecturing in classroom. To verify the effect of HoloBoard in actual classes, we designed a course and carried out a comparative study with 36 primary school students. The multimodal learning analytics results of the experiment demonstrated that students in the HoloBoard group were more engaged behaviorally, emotionally, and cognitively, indicating that immersive technology can benefit learning in lecture-based classes. We expect this work can contribute to the educational technology research community by providing a novel interactive system and a deep understanding of the immersive technology to support learning. 

\begin{acks}
This work was supported by the China National Key Research and Development Plan under Grant No. 2020YFF0305405. Our sincere thanks to all the participants of our experiment. 
\end{acks}

\bibliographystyle{ACM-Reference-Format}
\bibliography{HoloBoard}

\appendix

\section{Pre-/Post- Knowledge Test}

\subsection{Part One: Open-ended Questions }

\begin{enumerate}
\item Do you know any story about human exploring the space? You can tell me as detailed as possible.
\item Do you think why human explore the space? You can tell me as much as possible about the reasons that human explore the space.
\item Do you think why human explore the space? You can tell me as much as possible about the reasons that human explore the space.
\item Please describe the solar system.
\item Can you tell me what are the planets in the solar system? How can they be categorized? What are the characteristics of each of them?”
\item Do you think human can live freely in other planets? Why? What are the needed elements to make a planet habitable?
\end{enumerate}

\subsection{Part Two: Singe-choice Questions}
\begin{enumerate}
\item What is the largest planet in the solar system?

a. Earth  b. Jupiter  c. Sun  d. Uranus

\item In the solar system, what is the farthest planets from the sun?

a. Earth  b. Saturn  c. Mars  d. Neptune

\item What is the hottest planet in the solar system?

a. Venus  b. Saturn   c. Mars  d. Jupiter

\item What is the heaviest planet in the solar system? (the heaviest; the highest in mass)

a. Mercury  b. Earth  c. Mars  d. Jupiter
\end{enumerate}

\subsection{Part Three: Multiple-choice Questions}
\begin{enumerate}
\item In the following planets, what is/are the gas planet(s)?

a. Mercury  b. Venus  c. Neptune  d. Mars

\item What is/are the factor(s) effecting human moving to and living in other planet(s)? 

a. Water  b. Oxygen  c. Temperature  d. Distance
\end{enumerate}
\end{document}